\documentclass[aip,jcp,reprint,amsmath,amssymb,floatfix,citeautoscript,nofootinbib]{revtex4-2}
\usepackage{times}
\usepackage{mathptmx}
\DeclareMathAlphabet{\mathcal}{OMS}{cmsy}{m}{n} 
\usepackage[T1]{fontenc}
\usepackage{algorithm}
\usepackage{algpseudocode}
\usepackage{bm}
\usepackage{amsmath}
\usepackage{amssymb}
\usepackage{xcolor}
\usepackage{graphicx}
\usepackage{hyperref}
\hypersetup{
    colorlinks=true,
    linkcolor=blue,
    urlcolor=blue,
    citecolor=blue,
}
\usepackage{cleveref}
\usepackage{braket}
\usepackage[version=4]{mhchem}
\usepackage{multirow}

\definecolor{myred}{rgb}{0.8,0,0}

\newcommand{\myred}[1]{#1}
\newcommand{\myredd}[1]{#1}

\definecolor{mygreen}{rgb}{0,0.8,0}

\newcommand{\Hartree}{$E_{\mathrm{h}}$}

\begin{document}

    \title{Random phase approximation-based local natural orbital coupled cluster theory}

    \author{Ruiheng Song}
    \affiliation{Department of Chemistry and Biochemistry, University of Maryland, College Park, Maryland, 20742}

    \author{Xiliang Gong}
    \affiliation{Department of Chemistry and Biochemistry, University of Maryland, College Park, Maryland, 20742}

    \author{Aamy Bakry}
    \affiliation{Department of Chemistry and Biochemistry, University of Maryland, College Park, Maryland, 20742}

    \author{Hong-Zhou Ye}
    \email{hzye@umd.edu}
    \affiliation{Department of Chemistry and Biochemistry, University of Maryland, College Park, Maryland, 20742}
    \affiliation{Institute for Physical Science and Technology, University of Maryland, College Park, Maryland, 20742}
    \date{\today}

    \begin{abstract}
        Practical applications of fragment embedding and closely related local correlation methods depend critically on a judicious choice of low-level theory to define the local embedding subspace and to account for long-range electrostatic and correlation effects outside the embedding region.
        Second-order M{\o}ller-Plesset perturbation theory (MP2) is by far the most widely used correlated low-level theory; however, its applicability becomes questionable for systems in which MP2 is known to fail, either quantitatively or qualitatively.
        In this work, we present the random phase approximation (RPA) and the closely related second-order screened exchange (SOSEX) as promising alternatives to MP2 within the local natural orbital-based coupled-cluster (LNO-CC) framework.
        Through benchmark calculations on noncovalent molecular complexes and crystals, reaction barrier heights, and bulk metals, we demonstrate that RPA- and SOSEX-based LNO-CC at the LNO-CCSD(T) level closely match the performance of the corresponding MP2-based approach, while at the LNO-CCSD level they yield significantly faster convergence toward the canonical CCSD limit.
        The improvement is particularly pronounced for metallic systems as the thermodynamic limit is approached.
        These results highlight the critical role of the low-level theory in fragment embedding and local correlation methods and identify RPA as a compelling alternative to the commonly used MP2.
    \end{abstract}

    \maketitle

    \section{Introduction}

    Quantum embedding and fragment-based local correlation methods mitigate the steep computational cost of high-accuracy electronic structure theories such as coupled cluster~\cite{Bartlett07RMP} (CC) by applying these methods only to chemically meaningful local fragments~\cite{Sun16ACR,Herbert19JCP,Lee19ACR,Wasserman20IJQC}.
    Long-range electrostatic and correlation effects outside the fragments are instead treated with a low-level theory, rendering the overall approach computationally tractable.
    Representative examples include the use of one-body effective potentials in wavefunction-in-density functional theory embedding~\cite{Manby12JCTC,Daday14CPC,Jacob14WIRCMS,Libisch14ACR,Lee19ACR},
    mean-field or correlated bath orbitals in density matrix embedding theory~\cite{Knizia13JCTC,Wouters16JCTC,Cui20JCTC} (DMET) and its variants~\cite{Ye19JCTC,Cho25JPCA,Nusspickel22PRX,Shee24JCP,Hermes20JCTC,King25NCommun},
    and correlated natural orbitals in the local natural orbital~\cite{Rolik11JCP,Rolik13JCP,Nagy17JCP,Nagy19JCTC,Nagy24CS,Ye24JCTC} (LNO) extension of cluster-in-molecule~\cite{Li02JCC,Li04JCP,Li06JCP,Li09JCP,Wang19JCTC,Wang22JCTC} (CIM) methods.
    In all such approaches, the choice of the low-level theory plays a pivotal role, as it controls the rate of convergence with fragment size and thus largely determines the practical efficiency of the method.

    Second-order M{\o}ller-Plesset perturbation theory~\cite{Moller34PR} (MP2) is by far the most widely used correlated low-level theory in quantum embedding and local correlation methods.
    MP2-based natural orbitals and long-range correlation corrections have a long history in local CC theories~\cite{Neese09JCPa,Neese09JCPb,Hansen11JCP,Huntington12JCP,Liakos15JCTC,Werner11JCP}.
    More recently, MP2-derived natural orbitals have also been adopted as correlated bath states in DMET~\cite{Nusspickel22PRX,Nusspickel23JCTC} and dynamical mean-field theory~\cite{Li24PRL} (DMFT), where they alleviate the disentanglement problems associated with mean-field baths~\cite{Ye19JPCL} and enable an effective description of dynamical correlation in molecules and insulating solids.
    Despite these successes, the use of MP2 as a low-level theory becomes questionable in systems where MP2 is quantitatively inaccurate or even formally invalid.
    Two prominent examples, which are the focus of the present work, are noncovalently bound molecules and molecular solids---where MP2 significantly overestimates intermolecular interaction energies~\cite{Distasio07MP,Cybulski07JCP,Beran16CR,Liang23JPCL}---and bulk metals, for which the MP2 correlation energy diverges in the thermodynamic limit~\cite{Shepherd13PRL,Masios23PRL,Neufeld23PRL} (TDL).

    The random phase approximation~\cite{Bohm53PR,GellMann57PR,Furche01PRB} (RPA) provides a promising alternative to MP2 in both of these contexts~\cite{Ren12JMS,Chen17ARPC,Zhang21WIRCMS}.
    By resumming the direct particle-hole ring diagrams to infinite order, RPA incorporates dynamical screening of the bare Coulomb interaction,
    leading to improved descriptions of noncovalent interactions~\cite{Harl08PRB,Lu09PRL,DelBen13JCTC,Klimes16JCP,Yao21JPCL,Hellgren21PRR,Pham23JCP,Liang25JCTC}
    and metallic systems~\cite{Ren09PRB,Harl09PRL,Schimka10NM,Olsen11PRL,Rohlfin08PRL,Ma11PRB,Kim12PRB,Schmidt18JPCC,Sheldon24JCTC}.
    Within embedding frameworks, RPA has been most extensively explored in the form of constrained RPA (cRPA)~\cite{Aryasetiawan04PRB,Sasioglu11PRB,Chang24npjCM,Scott24PRL} for Hamiltonian downfolding, where the RPA polarizability is used to construct effective screened interactions for a chosen active space.
    More recently, Sch{\"a}fer and co-workers have demonstrated that RPA natural orbitals, obtained from first-order Green's functions~\cite{Ramberger19JCP}, can be used to accelerate embedding-based CC calculations~\cite{Schafer21JCPa,Schafer21JCPb}.
    Callahan and co-workers have demonstrated the use of RPA-based composite energy corrections to accelerate the convergence of CC calculations for uniform electron gas~\cite{Callahan21JCP}.

    In this work, we present an alternative route for incorporating RPA into fragment embedding and local correlation theories.
    Our approach is based on the direct ring coupled-cluster doubles (drCCD) formulation of RPA~\cite{Scuseria08JCP}, which provides direct access to RPA wavefunction amplitudes and can therefore be integrated seamlessly into existing MP2-based embedding and local correlation frameworks.
    We report an efficient implementation of RPA-based natural orbitals based on the factorized drCCD equation by He{\ss}elmann~\cite{Hebelmann12PRA} and K\'{a}llay~\cite{Kallay15JCP} with asymptotic cost scaling more favorable than that of conventional MP2-based approaches.
    The effectiveness of RPA as a low-level theory is demonstrated within the LNO-CC framework, originally developed by K\'{a}llay and co-workers for molecules~\cite{Rolik11JCP,Rolik13JCP,Nagy17JCP,Nagy19JCTC} and recently extended to both insulating and metallic solids by one of the authors~\cite{Ye23arXiv,Ye24FD,Ye24JCTC}.
    Our numerical results for noncovalently bound molecules, molecular solids, reaction barrier heights, and bulk metals indicate that RPA-derived LNOs are of comparable quality to their MP2-based counterparts, while composite energy corrections based on RPA and the closely related second-order screened exchange~\cite{Gruneis09JCP} (SOSEX) approximation lead to systematically faster convergence for LNO-CCSD with respect to LNO truncation than MP2-based corrections.
    This improvement is particularly pronounced for metallic systems as the TDL is approached using increasingly dense $k$-point meshes.
    Together, these results highlight the critical role of the low-level theory in fragment embedding and local correlation methods and suggest RPA as a compelling alternative to the commonly used MP2.

    The remainder of this paper is organized as follows.
    In \cref{sec:theory}, we briefly review the MP2-based LNO-CC formalism and introduce its RPA-based generalization, along with details of an efficient implementation for molecules and periodic solids.
    Computational details are given in \cref{sec:comp_details}.
    In \cref{sec:results_and_discussion}, we present benchmark results for two classes of systems in which MP2 is known to be problematic, namely noncovalently bound molecular systems and bulk metals, as well as for reaction barrier heights as a representative thermochemical application.
    Finally, \cref{sec:conclusion} summarizes our findings and outlines directions for future work.

    \section{Theory}
    \label{sec:theory}

    Throughout the paper, we assume a spin-restricted Hartree-Fock (HF) reference wavefunction with $N_{\textrm{occ}}$ occupied orbitals labelled by $i, j, k, \cdots$ and $N_{\textrm{vir}}$ virtual orbitals labelled by $a, b, c, \cdots$, which give rise to $N_{\textrm{MO}}$ molecular orbitals labelled by $p, q, r, \cdots$.
    The HF orbitals and their orbital energy are denoted by $\psi_{p}$ and $\varepsilon_{p}$, respectively.
    The Hamiltonian in the HF orbital basis reads
    \begin{equation}    \label{eq:Ham_HFbasis}
        \hat{H}
            = \sum_{\sigma} \sum_{pq}
            h_{pq} c_{p\sigma}^{\dagger} c_{q\sigma}
            + \frac{1}{2} \sum_{\sigma\sigma'} \sum_{pqrs}
            V_{pqrs} c^{\dagger}_{p\sigma} c^{\dagger}_{r\sigma'} c_{s\sigma'} c_{q\sigma}
    \end{equation}
    where $\sigma$ labels spin, $h_{pq}$ and $V_{pqrs}$ are the one- and two-electron integrals, with the latter in the $(11|22)$ notation.
    The formula presented below apply directly to molecules and periodic solids with a single $k$-point sampling the first Brillouin zone.
    Extension to periodic calculations with a uniform $k$-mesh is straightforward following Ref~\onlinecite{Ye24JCTC} by first transforming the HF orbitals into the corresponding supercell representation with a single $k$-point.

    \subsection{Local natural orbital-based coupled cluster (LNO-CC)}
    \label{subsec:lno_cc}

    In conventional MP2-based LNO-CC, the HF occupied orbitals are first localized to obtain an equal number of localized orbitals (LOs),
    \begin{equation}
        \phi_I
            = \sum_{i} \psi_i U_{iI},
    \end{equation}
    where the unitary matrix $U$ is determined by maximizing a chosen orbital locality metric (e.g.,~Pipek-Mezey~\cite{Pipek89JCP,Jonsson17JCTC}, Foster-Boys~\cite{Boys60RMP}, and the maximally localized Wannier functions~\cite{Marzari12RMP}).
    The LNOs associated with a given LO $\phi_I$ are obtained from the eigenvectors of the MP2 one-particle density matrix constructed for $\phi_I$,
    \begin{equation} \label{eq:lno_from_hfmo}
    \begin{split}
        \xi_{k}^{(I)}
            &= \sum_{i} \psi_i X^{(I)}_{ik}, \\
        \xi_{c}^{(I)}
            &= \sum_{a} \psi_a X^{(I)}_{ac},
    \end{split}
    \end{equation}
    where the occupied-occupied and virtual-virtual blocks of the density matrix are given by
    \begin{equation} \label{eq:mp2_dm}
    \begin{split}
        &D_{ij}^{(I)}
            = 2 \sum_{ab} t^{(1)*}_{iaIb} \left( 2 t^{(1)}_{jaIb} - t^{(1)}_{jbIa} \right)
            = \sum_{k} n^{(I)}_{k} X^{(I)}_{ik} X^{(I)*}_{jk} \\
        &D_{ab}^{(I)}
            = \sum_{jc} 2 \left(
                t^{(1)}_{Iajc} t^{(1)*}_{Ibjc}
                + t^{(1)}_{Icja} t^{(1)*}_{Icjb}
            \right) \\
        &\phantom{D_{ab}^{(I)} = \sum_{jc}}
            - \left(
                t^{(1)}_{Icja} t^{(1)*}_{Ibjc}
                + t^{(1)}_{Iajc} t^{(1)*}_{Icjb}
            \right)
            = \sum_{c} n^{(I)}_{c} X^{(I)}_{ac} X^{(I)*}_{bc}.
    \end{split}
    \end{equation}
    Here, $t_{Iajb}^{(1)} = \sum_{i} U_{iI} \, t_{iajb}^{(1)}$, and
    \begin{equation}    \label{eq:mp2_Tamp}
        t^{(1)}_{iajb}
            = \frac{V_{iajb}^*}{\varepsilon_i + \varepsilon_j - \varepsilon_a - \varepsilon_b}
    \end{equation}
    are the MP2 amplitudes and the asterisk symbol indicates complex conjugate.
    A commonly adopted approximation \myred{in LNO-CC and other local correlation theories} is to replace the full MP2 amplitudes in \cref{eq:mp2_dm} with their semi-canonical counterparts~\cite{Rolik11JCP,Neese09JCPa,Ye24JCTC}.
    In this work, however, we employ the full MP2 amplitudes for LNO construction in order to maintain consistency with the RPA-based LNO construction discussed in \cref{subsec:rpa_lno_cc}.

    Because the orbital rotations in \cref{eq:lno_from_hfmo} do not mix occupied and virtual subspaces, the resulting LNOs preserve the integer occupations of the underlying mean-field reference.
    Important occupied and virtual LNOs can therefore be selected independently to define a LO-specific local active space (LAS) $\mathcal{A}^{(I)}$ that is optimized for capturing the correlation energy associated with $\phi_I$.
    In this work, active LNOs are selected by thresholding the eigenvalues of \cref{eq:mp2_dm}, also known as LNO occupation numbers,
    \begin{equation} \label{eq:lno_selection}
    \begin{split}
        n_{i}^{(I)} &\geq \eta_{\mathrm{occ}}, \\
        n_{a}^{(I)} &\geq \eta_{\mathrm{vir}},
    \end{split}
    \end{equation}
    where $\eta_{\mathrm{occ}}$ and $\eta_{\mathrm{vir}}$ are user-defined thresholds.
    Tightening these thresholds systematically enlarges the LAS by including additional LNOs.

    The LNO-based LAS can be viewed as a correlated bath constructed for each individual LO, extending the mean-field bath concept employed in DMET~\cite{Knizia13JCTC,Wouters16JCTC} and related fragment embedding approaches~\cite{Ye19JCTC}.
    If the initial LOs span both the occupied space and part of the virtual space---as is common in fragment embedding methods~\cite{Wouters16JCTC,Ye19JPCL}---the LAS may be defined by combining a DMET-style mean-field bath derived from Schmidt decomposition~\cite{Ekert95AJP} with truncated LNOs obtained from the remaining orbital space~\cite{Ye24JCTC}.
    This generalization enables the use of more flexible LO types, such as intrinsic atomic orbitals~\cite{Knizia13JCTC}, thereby facilitating the application of LNO-CC to systems with small energy gaps (e.g., metals), for which localization of the occupied orbitals alone can be numerically challenging~\cite{Marzari12RMP,Cornean19AHP}.

    LNO-CC exploits the LNO-based LAS to approximate the canonical CC correlation energy.
    In this work, we focus primarily on LNO-CCSD and LNO-CCSD(T).
    The CCSD correlation energy is approximated as
    \begin{equation} \label{eq:Ec_LNOCCSD_sum_I}
        E_{\mathrm{c,CCSD}}
            \approx E_{\mathrm{c,LNO\text{-}CCSD}}
            = \sum_{I} E_{\mathrm{c,LNO\text{-}CCSD}}^{(I)}
            + \Delta E_{\mathrm{c,MP2}}^{\mathrm{ext}},
    \end{equation}
    where the first term collects CCSD correlation energies evaluated within the LAS of each LO, and the second term is an external correction that approximately accounts for correlation energy arising from orbitals outside the LAS at the MP2 level.
    To evaluate both contributions, a local Hamiltonian is constructed for each LO by projecting the full Hamiltonian $\hat{H}$ onto the corresponding LAS,
    \begin{equation}
        \hat{H}^{(I)}
            = \sum_{\sigma} \sum_{pq \in \mathcal{A}^{(I)}}
            h_{pq}^{(I)} c_{p\sigma}^{\dagger} c_{q\sigma}
            + \frac{1}{2} \sum_{\sigma\sigma'} \sum_{pqrs \in \mathcal{A}^{(I)}}
            V_{pqrs}
            c^{\dagger}_{p\sigma} c^{\dagger}_{r\sigma'}
            c_{s\sigma'} c_{q\sigma},
    \end{equation}
    where \(
        h_{pq}^{(I)}
            = h_{pq} +
                \sum_{i \notin \mathcal{A}^{(I)}}
            \left( 2 V_{pqii} - V_{piiq} \right)
    \).
    Solving the CCSD equations for each $\hat{H}^{(I)}$ yields the local CCSD amplitudes $t_{ia}^{(I)}$ and $t_{iajb}^{(I)}$ within the LAS.
    These amplitudes define the LO-resolved CCSD correlation energy,
    \begin{equation} \label{eq:Ec_LNOCCSD_I}
        E_{\mathrm{c,LNO\text{-}CCSD}}^{(I)}
            = \sum_{jab \in \mathcal{A}^{(I)}}
            \tau^{(I)}_{Iajb} \left( 2 V_{Iajb} - V_{Ibja} \right),
    \end{equation}
    where $\tau^{(I)}_{Iajb} = t^{(I)}_{Iajb} + t^{(I)}_{Ia} t^{(I)}_{jb}$ \myred{and the partially transformed CCSD amplitudes are defined as,
    \begin{equation}    \label{eq:t_amp_LO_transform}
    \begin{split}
        t_{Ia}^{(I)}
            &= \sum_{i} t_{ia}^{(I)} U_{iI},    \\
        t_{Iajb}^{(I)}
            &= \sum_{i} t_{iajb}^{(I)} U_{iI},
    \end{split}
    \end{equation}}%
    The MP2-based external correction is evaluated as the difference between the canonical MP2 correlation energy and its LNO-truncated counterpart,
    \begin{equation} \label{eq:Ec_ext_MP2}
        \Delta E_{\mathrm{c,MP2}}^{\mathrm{ext}}
            = E_{\mathrm{c,MP2}} - \sum_{I} E_{\mathrm{c,LNO\text{-}MP2}}^{(I)},
    \end{equation}
    where
    \begin{equation} \label{eq:Ec_LNOMP2_I}
        E_{\mathrm{c,LNO\text{-}MP2}}^{(I)}
            = \sum_{jab \in \mathcal{A}^{(I)}}
            t^{(1)(I)}_{Iajb} \left( 2 V_{Iajb} - V_{Ibja} \right),
    \end{equation}
    \myred{with $t^{(1)(I)}_{Iajb}$ obtained similarly from \cref{eq:t_amp_LO_transform} using the canonical MP2 amplitudes in subspace $\mathcal{A}^{(I)}$.}
    The CCSD(T) correlation energy can be approximated in an analogous fashion; detailed expressions may be found in Ref~\onlinecite{Nagy17JCP,Ye24JCTC}.

    Because each LAS consists of LNOs localized around a given LO, the MP2-based LNO-CC energy [\cref{eq:Ec_LNOCCSD_sum_I}] is accurate when the long-range component of the correlation energy is well described at the MP2 level.
    A key practical advantage of the LNO framework is its systematic improvability: tightening the truncation thresholds in \cref{eq:lno_selection} enlarges the LAS and reduces the approximation error.
    In the limit of vanishing thresholds, the LAS spans the full orbital space, and LNO-CCSD and LNO-CCSD(T) recover their canonical counterparts.
    For fixed thresholds, the LAS size has been shown to be asymptotically independent of system size, rendering LNO-CCSD and LNO-CCSD(T) linear-scaling methods~\cite{Rolik13JCP,Nagy17JCP}.
    In practice, however, the usefulness of the LNO approximation depends critically on the rate at which the error decays with LAS size, which is governed by the quality of the underlying low-level theory.
    In the following section, we introduce RPA as an alternative low-level theory for LNO-CC.

    \subsection{RPA-based LNO-CC}
    \label{subsec:rpa_lno_cc}

    As discussed in \cref{subsec:lno_cc}, MP2 serves two distinct roles in conventional LNO-CC: (i) the construction of LNOs [\cref{eq:lno_from_hfmo,eq:mp2_dm}] and (ii) the evaluation of the external energy correction [\cref{eq:Ec_ext_MP2}].
    The CC-based formulation of RPA~\cite{Scuseria08JCP,Henderson10MP,Scuseria13JCP,Peng13JCP,Rekkedal13JCP,Liang25JCTC} provides a natural alternative to MP2 for both tasks, leading to what we refer to as RPA-based LNO-CC.
    Within the CC framework, RPA is equivalent to the direct ring CCD (drCCD) approximation~\cite{Scuseria08JCP}, which retains only the direct particle-hole ring diagrams of CCSD.
    The RPA correlation energy is given by
    \begin{equation} \label{eq:Ec_drCCD}
        E_{\mathrm{c,RPA}}
            = E_{\mathrm{c,drCCD}}
            = 2 \sum_{ijab} T_{iajb} V_{iajb},
    \end{equation}
    where the drCCD amplitudes $T_{iajb}$ satisfy the nonlinear amplitude equation
    \begin{equation} \label{eq:drCCD_Tamp_eqn}
        B^* + \Delta \circ T + 2 L^* T + 2 T L + 4 T B T
            = 0.
    \end{equation}
    Here, $B_{iajb} = V_{iajb}$, $L_{iajb} = V_{iabj}$, and $\Delta_{iajb} = \varepsilon_a + \varepsilon_b - \varepsilon_i - \varepsilon_j$.
    The symbol ``$\circ$'' denotes the Hadamard (element-wise) product, and matrix multiplication is defined as $(AB)_{iajb} = \sum_{kc} A_{iakc} B_{kcjb}$.
    Because the nonlinear drCCD equations can admit multiple solutions~\cite{Henderson10MP,Rekkedal13JCP,Song25JCP}, we employ the level-shift preconditioner described in Ref~\onlinecite{Song25JCP} to ensure robust convergence to the physical solution.

    The extension of LNO-CC from MP2 to RPA is straightforward because of the formal similarity between the MP2 and drCCD amplitudes.
    Specifically, we construct RPA-based LNOs from \myredd{MP2-inspired} one-particle density matrices obtained from \cref{eq:mp2_dm} by replacing the MP2 amplitudes in \cref{eq:mp2_Tamp} with the drCCD amplitudes $T_{iajb}$.
    \myredd{This choice preserves the computational efficiency of the existing MP2-based LNO generation procedure while allowing us to directly assess the effect of the $T$ amplitudes on the quality of the resulting LNOs.
    More rigorous definitions of RPA one-particle density matrices, for example those derived from a Lagrangian or analytic-gradient formalism~\cite{Rekkedal13JCP,Ramberger17PRL}, will be explored in future work.}
    The RPA-based external energy correction is defined analogously to \cref{eq:Ec_ext_MP2},
    \begin{equation} \label{eq:Ec_ext_RPA}
        \Delta E_{\mathrm{c,RPA}}^{\mathrm{ext}}
            = E_{\mathrm{c,RPA}} - \sum_{I} E_{\mathrm{c,LNO\text{-}RPA}}^{(I)},
    \end{equation}
    where the LO-resolved RPA correlation energy follows directly from \cref{eq:Ec_drCCD},
    \begin{equation} \label{eq:Ec_LNORPA_I}
        E_{\mathrm{c,LNO\text{-}RPA}}^{(I)}
            = 2 \sum_{jab \in \mathcal{A}^{(I)}} T^{(I)}_{Iajb} V_{Iajb}.
    \end{equation}

    A well-known limitation of RPA is the neglect of exchange diagrams, which typically leads to an overestimation of the correlation energy due to self-interaction.
    Including exchange contributions in the evaluation of the drCCD correlation energy [\cref{eq:Ec_drCCD}] leads to the RPA+SOSEX method~\cite{Gruneis09JCP} (referred to as SOSEX for simplicity hereafter),
    \begin{equation} \label{eq:Ec_SOSEX}
        E_{\mathrm{c,SOSEX}}
            = \sum_{ijab} T_{iajb} \left( 2 V_{iajb} - V_{ibja} \right),
    \end{equation}
    which removes the one-electron self-interaction error \myred{inherent in the RPA energy}~\cite{Gruneis09JCP,Hebelmann12PRA}.
    Employing SOSEX to evaluate the external energy correction leads to what we term the SOSEX-based LNO-CC approach,
    \begin{equation} \label{eq:Ec_ext_SOSEX}
        \Delta E_{\mathrm{c,SOSEX}}^{\mathrm{ext}}
            = E_{\mathrm{c,SOSEX}}
            - \sum_{I} E_{\mathrm{c,LNO\text{-}SOSEX}}^{(I)},
    \end{equation}
    where the LO-resolved SOSEX correlation energy is given by
    \begin{equation}
        E_{\mathrm{c,LNO\text{-}SOSEX}}^{(I)}
            = \sum_{jab \in \mathcal{A}^{(I)}}
            T^{(I)}_{Iajb} \left( 2 V_{Iajb} - V_{Ibja} \right).
    \end{equation}
    More generally, the RPA-based LNO-CC framework can be straightforwardly extended to incorporate other beyond-RPA correlation methods formulated within the CC framework~\cite{Hebelmann12PRA,Shepherd14JCP,Cieslinski23JCTC}, which we leave for future work.

    \myred{In passing, we note that one may neglect the exchange terms in \cref{eq:mp2_dm} to define what we term the direct density matrices, whose diagonalization yields the corresponding direct LNOs (dLNOs),
    \begin{equation} \label{eq:rpa_direct_dm}
    \begin{split}
        D_{ij}^{\mathrm{d}(I)}
            &= 4 \sum_{ab} T^*_{iaIb} T_{jaIb}
            = \sum_{k} n^{\mathrm{d}(I)}_{k}
            X^{\mathrm{d}(I)}_{ik} X^{\mathrm{d}(I)*}_{jk}, \\
        D_{ab}^{\mathrm{d}(I)}
            &= 2 \sum_{jc}
            \left(
                T_{Iajc} T^*_{Ibjc}
                + T_{Icja} T^*_{Icjb}
            \right)
            = \sum_{c} n^{\mathrm{d}(I)}_{c}
            X^{\mathrm{d}(I)}_{ac} X^{\mathrm{d}(I)*}_{bc}.
    \end{split}
    \end{equation}
    For RPA, one may view \cref{eq:rpa_direct_dm} as conceptually more natural than the MP2-like definition in \cref{eq:mp2_dm}, because RPA itself lacks exchange contributions.
    In practice, however, our numerical tests indicate that LNOs and dLNOs yield very similar performance in LNO-CC calculations at comparable LAS sizes (fig.~S1).
    We therefore do not pursue this variant further here and instead adopt the regular LNO definition based on \cref{eq:mp2_dm} throughout the remainder of this work.}

    \subsection{Efficient implementation of RPA-based LNO-CC}

    The drCCD equation (\ref{eq:drCCD_Tamp_eqn}) can be solved efficiently using the quartic-scaling algorithms originally developed by He{\ss}elmann~\cite{Hebelmann12PRA} and K\'{a}llay~\cite{Kallay15JCP}.
    In these approaches, the four-index amplitudes $T_{iajb}$ are factorized in terms of three-index intermediates,
    \begin{equation} \label{eq:drCCD_T_U}
        T_{iajb}
            = - \sum_{w}^{N_{\mathrm{CD}}} \tau_{iaw}
            \left(
                \sum_{P}^{N_{\mathrm{aux}}} U_{iaP} U_{jbP}
            \right)
            \tau_{jbw},
    \end{equation}
    where $\tau_{iaw}$ are Cholesky factors of the orbital-energy denominator~\cite{Koch00JCP,Aquilante11Chapter},
    \begin{equation} \label{eq:cd_denom}
        (\Delta_{iajb})^{-1}
            \approx \sum_{w}^{N_{\mathrm{CD}}} \tau_{iaw} \tau_{jbw},
    \end{equation}
    and $P$ labels the auxiliary basis functions used in density fitting (DF) of the electron repulsion integrals~\cite{Whitten73JCP,Feyereisen93CPL,Dunlap00PCCP},
    \begin{equation} \label{eq:df_eris}
    \begin{split}
        B_{iajb}
            &= \sum_{P}^{N_{\mathrm{aux}}} J_{iaP} J_{jbP}, \\
        L_{iajb}
            &= \sum_{P}^{N_{\mathrm{aux}}} J_{iaP} J^*_{jbP}.
    \end{split}
    \end{equation}
    The three-index amplitudes $U_{iaP}$ satisfy the factorized drCCD equation~\cite{Hebelmann12PRA,Kallay15JCP},
    \begin{equation} \label{eq:drCCD_Uamp_eqn}
        U_{iaP}
            = J^*_{iaP}
            - \sum_{w}^{N_{\mathrm{CD}}} \tau_{iaw}
            \left(
                \sum_{Q}^{N_{\mathrm{aux}}} U_{iaQ} U_{jbQ}
            \right)
            J_{jbP}.
    \end{equation}
    \Cref{eq:drCCD_Uamp_eqn} is independent of the four-index $T$ amplitudes and therefore requires only $O(N^3)$ memory, comparable to DF-MP2.
    The computational cost of solving \cref{eq:drCCD_Uamp_eqn} scales as $O(N^4)$, which is asymptotically more favorable than DF-MP2.

    Once the $U$ amplitudes are obtained, the total RPA correlation energy can be evaluated without explicitly forming $T$,
    \begin{equation}
        E_{\mathrm{c,RPA}}
            = 2 \sum_{ia} \sum_{P}^{N_{\mathrm{aux}}}
            \left( J^*_{iaP} - U_{iaP} \right) J_{iaP}.
    \end{equation}
    Similarly, the LO-resolved RPA correlation energy required for the external correction in \cref{eq:Ec_ext_RPA} can be evaluated using the local $U$ amplitudes restricted to the LAS,
    \begin{equation}
        E_{\mathrm{c,LNO\text{-}RPA}}^{(I)}
            = 2 \sum_{a \in \mathcal{A}^{(I)}} \sum_{P}^{N_{\mathrm{aux}}}
            \left(
                J^*_{IaP} - U^{(I)}_{IaP}
            \right) J_{IaP}.
    \end{equation}
    The evaluation of drCCD density matrices [\cref{eq:mp2_dm}] required for constructing RPA-based LNOs and the calculation of the SOSEX external correction [\cref{eq:Ec_ext_SOSEX}] both necessitate reconstructing the $T$ amplitudes from the factorized representation in \cref{eq:drCCD_T_U}.
    For each LO, the associated CPU and memory costs scale as $O(N^4)$ and $O(N^3)$, respectively, with the peak memory requirement further reducible to $O(N^2)$ by constructing $T$ in batches.

    The quartic-scaling drCCD algorithm was originally developed for molecular systems~\cite{Hebelmann12PRA,Kallay15JCP}.
    Here, we extend this framework to periodic solids with explicit $k$-point sampling, enabling periodic RPA-based LNO-CC calculations.
    The computational cost of the periodic drCCD implementation scales as $O(N_k^2 n^4)$, which is asymptotically more favorable than the $O(N_k^3 n^5)$ scaling of periodic DF-MP2, \myred{where $N_k$ denotes the number of $k$-points sampling the Brillouin zone}.
    The corresponding memory cost scales as $O(N_k^2 n^3)$ for storing the $U$ amplitudes, which is modest and comparable to that of DF-MP2.
    Further implementation details of the quartic-scaling periodic drCCD algorithm will be reported elsewhere~\cite{GongInprep}.

    \section{Computational details}
    \label{sec:comp_details}

    We have implemented RPA-based LNO-CCSD and LNO-CCSD(T) for both molecular and periodic systems in a developer version of the \textsc{PySCF} package~\cite{Sun18WIRCMS,Sun20JCP}, which employs \textsc{libcint}~\cite{Sun15JCC} for Gaussian integral evaluation.
    The number of Cholesky vectors $N_{\mathrm{CD}}$ retained in the factorized drCCD equation (\ref{eq:drCCD_Uamp_eqn}) is chosen sufficiently large to ensure a negligible fitting error,
    \begin{equation}
        \max \left|
            \Delta_{\min}
            \left[
                (\Delta_{iajb})^{-1}
                - \sum_{w}^{N_{\mathrm{CD}}} \tau_{iaw} \tau_{jbw}
            \right]
        \right|
        < \tau_{\mathrm{CD}},
    \end{equation}
    where $\Delta_{\min} = \min \Delta_{iajb}$ normalizes the fitting error to the interval $(0,1]$.
    Throughout this work, we adopt $\tau_{\mathrm{CD}} = 10^{-5}$, which typically results in $N_{\mathrm{CD}} = 5$--$10$ and yields both RPA and SOSEX correlation energy errors below $10^{-7}$~\Hartree{} per atom for both molecules and solids.
    \myred{Our current drCCD-based RPA implementation is approximately $5$--$10$ times slower than the MP2 implementation in PySCF for the systems considered in this work.
    Its efficiency can, however, be improved in future work by incorporating recently developed linear-scaling drCCD algorithms~\cite{Liang25JCTC}.
    We also note that, for most LNO-CC calculations presented in this work, the overall computational cost is dominated by the fragment CC calculations, as illustrated in Table S1.
    }

    For periodic calculations, the integrable divergence of the HF exchange energy is treated using the probe-charge Ewald method~\cite{Paier05JCP}, which amounts to shifting the occupied orbital energies downwards by the Madelung constant~\cite{Broqvist09PRB}.
    Following previous work to achieve optimal finite-size error decay~\cite{Ye24JCTC,Xing24PRX}, the Madelung-shifted orbital energies are used in periodic MP2, RPA, and the perturbative (T) correction of CCSD(T), whereas the unshifted orbital energies are employed in the periodic CCSD equations.
    The electron repulsion integrals in periodic systems are handled using the range-separated DF scheme~\cite{Ye21JCPa,Ye21JCPb}.

    For the noncovalence molecular complexes and solids studied in \cref{subsec:non_covalent_molecules} and the reaction barrier heights studied in \cref{subsec:barrier_height}, we employ standard all-electron cc-pVTZ~\cite{Dunning89JCP} or aug-cc-pVTZ~\cite{Kendall92JCP} basis sets together with the corresponding resolution-of-identity fitting basis sets~\cite{Weigend02PCCP} for DF.
    The core electrons of all non-hydrogen atoms are frozen in correlated calculations.
    For bulk lithium studied in \cref{subsec:bulk_metals}, we use cc-pVDZ and TZ basis sets~\cite{Ye22JCTC} optimized for the small-core Goedecker--Teter--Hutter (GTH)-HF pseudopotentials~\cite{Goedecker96PRB,Hartwigsen98PRB,HutterPP}, without freezing any electrons.
    For bulk copper, also discussed in \cref{subsec:bulk_metals}, we reoptimized the all-electron cc-pVTZ basis sets following the procedure described in Ref~\onlinecite{Ye22JCTC} to mitigate severe linear dependencies present in the standard cc-pVTZ sets~\cite{Balabanov05JCP}.
    The [Ar] core electrons are frozen for Cu.
    For both metallic systems, density fitting is performed using fitting basis sets optimized according to Ref~\onlinecite{Weigend02PCCP}.

    In all LNO-CC calculations, we fix the ratio between the occupied and virtual LNO truncation thresholds, $\gamma = \eta_{\mathrm{occ}} / \eta_{\mathrm{vir}}$, and vary a single parameter---chosen as $\eta_{\mathrm{vir}}$---to control the size of the LAS for approaching the canonical CC limit.
    For molecular and molecular-crystal systems studied in \cref{subsec:non_covalent_molecules,subsec:barrier_height}, we follow Ref~\onlinecite{Ye24JCTC} and use $\gamma = 10$ together with Pipek-Mezey LOs~\cite{Pipek89JCP,Jonsson17JCTC} constructed using the meta-L\"{o}wdin projector~\cite{Sun14JCTC}.
    For bulk metals studied in \cref{subsec:bulk_metals}, we adopt $\gamma = 1$ and employ intrinsic atomic orbitals~\cite{Knizia13JCTC} as the LOs.

    \section{Results and discussion}
    \label{sec:results_and_discussion}

    \subsection{Noncovalent molecular complexes and solids}
    \label{subsec:non_covalent_molecules}

\begin{figure}[!t]
        \centering
        \includegraphics[width=3in]{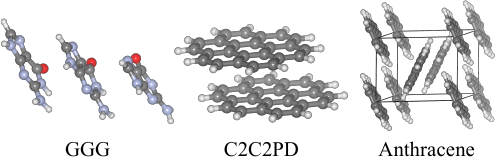}
        \caption{Structure of the guanine trimer (GGG), the coronene dimer in the parallel-displaced configuration (C2C2PD), and the anthracene crystal.
        Color scheme: C (gray), H (white), N (light blue), O (red).
        }
        \label{fig:mol_struct}
    \end{figure}

    \myred{Our first test set includes two noncovalent molecular complexes---the guanine trimer (GGG) and the coronene dimer in the parallel-displaced configuration (C2C2PD)---taken from the L7 dataset~\cite{Sedlak13JCTC}, as well as a molecular solid, the anthracene crystal, taken from the X23 dataset~\cite{Dolgonos19PCCP}.
    Their structures are visualized in \cref{fig:mol_struct}.
    The binding energy $E_{\mathrm{bind}}$ of both molecular complexes has been extensively studied using a range of correlated wavefunction methods~\cite{Sedlak13JCTC,AlHamdani21NCommun,Schafer25NCommun,Liang25JCTC}.
    High-level diffusion Monte Carlo~\cite{AlHamdani21NCommun} and CCSD(cT)~\cite{Schafer25NCommun} (CCSD with a modified perturbative triples correction~\cite{Masios23PRL}) benchmarks reveal a systematic overestimation of $E_{\mathrm{bind}}$ by MP2 and an underestimation by RPA based on a HF reference.
    Interestingly, despite the overbinding at the MP2 level, MP2-based LNO-CCSD(T) has been shown to yield reasonably accurate results for both molecular complexes~\cite{AlHamdani21NCommun}.
    The lattice energy $E_{\mathrm{lat}}$ of molecular solids, including the anthracene crystal, has also been the subject of several recent computational studies employing correlated wavefunction methods~\cite{DelBen13JCTC,Klimes16JCP,Zen18PNAS,Pham23JCP,DellaPia24PRL,Liang23JPCL}, which likewise suggest a systematic overestimation of $E_{\mathrm{lat}}$ by MP2~\cite{Liang23JPCL}.
    To the best of our knowledge, local CC calculations for molecular solids under periodic boundary conditions have not yet been reported.}

\begin{figure*}[!t]
        \centering
        \includegraphics[width=6in]{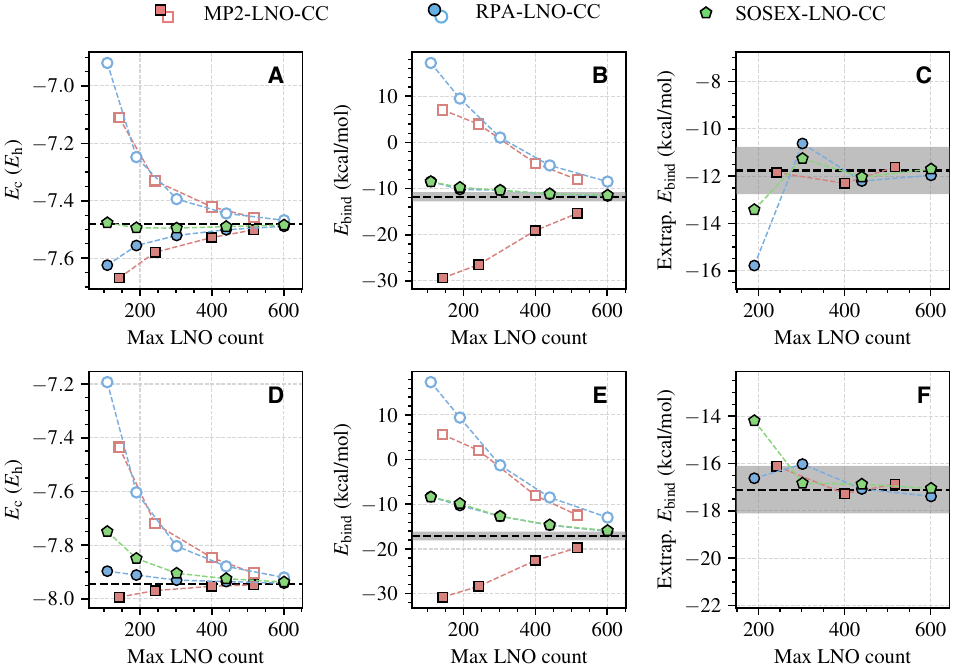}
        \caption{(A--C) Convergence of MP2-, RPA-, and SOSEX-based LNO-CCSD correlation energies (A), binding energies (B), and extrapolated binding energies (C) for the coronene dimer in the parallel-displaced configuration (C2C2PD) as a function of the maximum LAS size, measured by the number of active LNOs.
        (D--F) Corresponding results for LNO-CCSD(T).
        In (A,B,D,E), open and filled symbols denote LNO-CCSD or LNO-CCSD(T) results obtained without and with the external correction $\Delta E_{\mathrm{c}}^{\mathrm{ext}}$, respectively.
        In (C,F), extrapolated values are obtained using \cref{eq:Ec_extrap_ext} based on two adjacent LNO thresholds; the abscissa indicates the LAS size associated with the tighter threshold.
        The best estimate of the canonical CCSD or CCSD(T) result, obtained by averaging the best extrapolations from both MP2- and RPA-based LNO constructions, is shown as a horizontal dashed line in each panel.
        The gray shaded region in each panel represents an error window of $\pm 1$~kcal/mol relative to the corresponding reference value.
        All calculations were performed using the cc-pVTZ basis sets with the frozen core approximation.
        The total orbital count (excluding frozen orbitals) is $1776$.
        The binding energies are counterpoise-corrected.
        }
        \label{fig:c2c2pd_conv}
    \end{figure*}

    For each system, we performed LNO-CCSD and LNO-CCSD(T) calculations using a sequence of progressively tighter LNO truncation thresholds ($3\times10^{-5}, 10^{-5}, 3\times10^{-6}, \cdots$), and monitored the convergence of both the correlation energy and the binding or lattice energy as a function of the maximum LAS size, measured by the number of active LNOs.
    For a given truncation threshold, RPA-based LNOs yield a slightly smaller LAS size than their MP2-based counterparts, reflecting the screening effect in RPA, which regularizes the magnitude of the drCCD $T$ amplitudes in the presence of small orbital energy gaps.
    The results for C2C2PD obtained with MP2- and RPA-based LNO constructions and external corrections are summarized in \cref{fig:c2c2pd_conv}.

    \myred{We first examine the effect of the low-level theory on the LNO-CCSD correlation energy, as shown in \cref{fig:c2c2pd_conv}A.
    When the external energy correction $\Delta E_{\mathrm{c}}^{\mathrm{ext}}$ is not included, the MP2- and RPA-based LNO-CCSD correlation energies exhibit very similar convergence behavior with increasing LAS size, indicating that the two LNO constructions are of comparable quality for these noncovalently bound systems.
    A clear difference emerges, however, when external corrections are applied.
    MP2-based corrections [\cref{eq:Ec_ext_MP2}] lead to a systematic overestimation of the CCSD correlation energy.
    RPA-based corrections [\cref{eq:Ec_ext_RPA}] consistently mitigate the overestimation introduced by MP2, resulting in smaller energy errors at comparable LAS sizes.
    The fastest convergence is obtained with SOSEX-based corrections [\cref{eq:Ec_ext_SOSEX}], which already reach near convergence with only $\sim 100$ active LNOs, whereas the RPA- and MP2-based corrections require more than $400$ active LNOs to achieve a comparable accuracy.}

    In \cref{fig:c2c2pd_conv}B, we show the corresponding convergence of the counterpoise-corrected binding energy $E_{\mathrm{bind}}$ of C2C2PD, including the HF contribution.
    Overall, the convergence of $E_{\mathrm{bind}}$ closely mirrors that of $E_{\mathrm{c}}$ shown in \cref{fig:c2c2pd_conv}A.
    \myred{One important difference is that the RPA- and SOSEX-based corrections now lead to nearly identical results, with both approaches converging to within chemical accuracy using approximately $400$ active LNOs, whereas the MP2-based correction still leaves a sizable error of approximately $-5$~kcal/mol at a comparable LAS size.}
    To further accelerate convergence, we perform a linear extrapolation of $E_{\mathrm{bind}}$ to the limit $\Delta E_{\mathrm{c}}^{\mathrm{ext}} = 0$ using the following two-point formula~\cite{Nagy21MP}:
    \begin{equation}
    \label{eq:Ec_extrap_ext}
        E_{\textrm{c,CC}}
            \approx \frac{
                \Delta E_{\textrm{c}}^{\textrm{ext}}(\eta_2) E_{\textrm{c,LNO-CC}}(\eta_1) -
                \Delta E_{\textrm{c}}^{\textrm{ext}}(\eta_1) E_{\textrm{c,LNO-CC}}(\eta_2)
            }{
                \Delta E_{\textrm{c}}^{\textrm{ext}}(\eta_2) -
                \Delta E_{\textrm{c}}^{\textrm{ext}}(\eta_1)
            }
    \end{equation}
    where $\eta_1$ and $\eta_2$ denote two adjacent LNO truncation thresholds separated by approximately half a logarithmic unit.
    As demonstrated in \cref{fig:c2c2pd_conv}C, this extrapolation scheme is highly effective for both LNO constructions, yielding chemically accurate $E_{\mathrm{bind}}$ with as few as roughly $300$ active LNOs, which is only a small fraction of the total number of orbitals in the system (1776 orbitals).

    \myred{The LNO-CCSD(T) results for C2C2PD corresponding to the LNO-CCSD results discussed above are presented in \cref{fig:c2c2pd_conv}D--F.
    When external energy corrections are not included, the overall convergence behavior of LNO-CCSD(T) closely parallels that of LNO-CCSD for both $E_{\mathrm{c}}$ and $E_{\mathrm{bind}}$, further confirming the comparable quality of the MP2- and RPA-based LNO constructions.
    For a given LNO truncation threshold, however, the error in the uncorrected LNO-CCSD(T) $E_{\mathrm{c}}$ is slightly larger than the corresponding LNO-CCSD error, as can be seen by comparing \cref{fig:c2c2pd_conv}A and C.
    This additional error propagates into the externally corrected LNO-CCSD(T) values of $E_{\mathrm{c}}$ and $E_{\mathrm{bind}}$, shifting them toward more positive values relative to the corresponding externally corrected LNO-CCSD results, regardless of the type of external correction employed.
    This shift generally improves the accuracy of the MP2-corrected LNO-CCSD(T) results, while deteriorating that of the SOSEX-corrected ones.
    For the RPA-based correction, the convergence of $E_{\mathrm{c}}$ is improved, whereas that of $E_{\mathrm{bind}}$ remains similar to the SOSEX-corrected results and is therefore slightly worse than the LNO-CCSD counterpart.
    Nevertheless, the extrapolation based on \cref{eq:Ec_extrap_ext} remains robust and yields chemically accurate $E_{\mathrm{bind}}$ for all three types of external corrections using about $300$ active LNOs.}

    \begin{figure}[!t]
        \centering
        \includegraphics[width=3in]{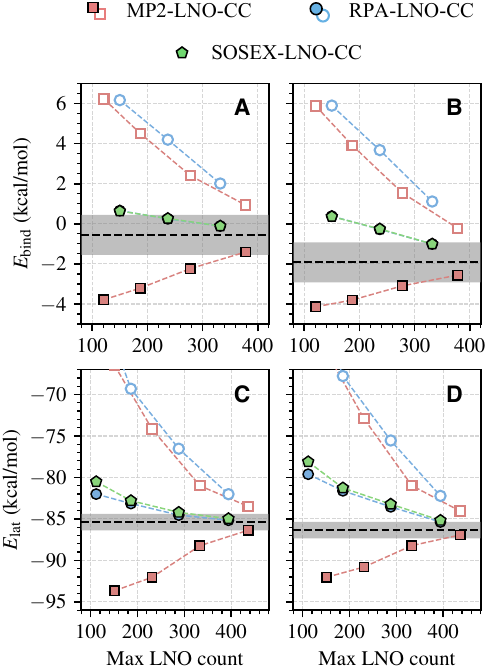}
        \caption{\myred{(A,B) Convergence of MP2-, RPA-, and SOSEX-based LNO-CCSD (A) and LNO-CCSD(T) (B) binding energy for the guanine trimer (GGG) as a function of the maximum LAS size, measured by the number of active LNOs.
        (C,D) Corresponding results for the lattice energy of the anthracene molecular crystal evaluated with $\Gamma$-point Brillouin zone sampling.
        In each panel, the best estimate of the canonical CCSD or CCSD(T) result, obtained by averaging the best extrapolations from the MP2-, RPA-, and SOSEX-based LNO-CC results, is shown as a horizontal dashed line.
        The gray shaded region denotes an error window of $\pm 1$~kcal/mol relative to the corresponding reference value.
        The aug-cc-pVTZ basis sets are used for GGG, and the cc-pVTZ basis sets are used for the anthracene crystal.
        The frozen-core approximation is used in all calculations.
        The total number of orbitals (excluding frozen orbitals) is $1863$ for GGG and $1120$ per unit cell for the anthracene crystal.}
        }
        \label{fig:ggg_anthra_conv}
    \end{figure}

    \myred{The convergence behavior observed for C2C2PD is further supported by analogous benchmark results for GGG and the anthracene crystal.
    As shown in \cref{fig:ggg_anthra_conv} for $E_{\mathrm{bind}}$ and $E_{\mathrm{lat}}$, and in figs.~S2 and S3 for $E_{\mathrm{c}}$ and the extrapolated results, these two systems exhibit trends very similar to those of C2C2PD.
    In particular, \cref{fig:ggg_anthra_conv} shows that LNO-CCSD converges most rapidly when combined with RPA- or SOSEX-based corrections, whereas LNO-CCSD(T) benefits slightly more from the MP2-based correction.
    This difference is likely related to the distinct physical content of the external treatments.
    Because RPA provides a more balanced description of long-range correlation than MP2, the RPA- and SOSEX-based external corrections are expected to resemble the missing CCSD correlation outside the LAS more closely.
    For LNO-CCSD(T), this balance shifts somewhat in favor of MP2, likely because the perturbative character of MP2 is more consistent with that of the perturbative triples correction in CCSD(T), leading to a more similar residual error structure outside the LAS.}

    \subsection{Reaction barrier heights}
    \label{subsec:barrier_height}

    \begin{figure}[!b]
        \centering
        \includegraphics[width=3in]{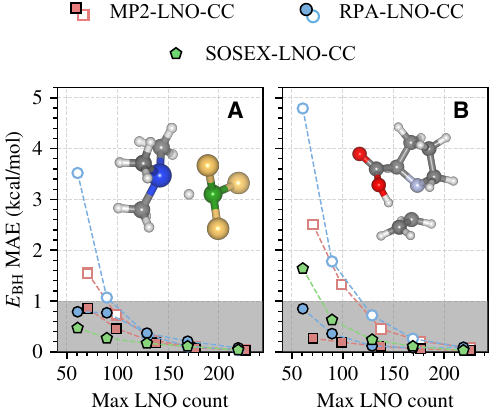}
        \caption{\myred{Convergence of the mean absolute error (MAE) in MP2-, RPA-, and SOSEX-based (A) LNO-CCSD and (B) LNO-CCSD(T) barrier heights ($E_{\text{BH}}$) for the BHDIV10 dataset~\cite{Goerigk17PCCP} as a function of the maximum LAS size averaged over all 10 systems.
        Open and filled symbols denote LNO-CC results obtained without and with the external energy correction, respectively.
        The uncorrected SOSEX-based LNO-CC results are identical to the uncorrected RPA-based results and are therefore not shown.
        Errors are evaluated relative to the canonical CCSD and CCSD(T) barrier heights.
        All calculations employ all-electron aug-cc-pVTZ basis sets, with the core electrons frozen for non-hydrogen atoms.
        The transition state structure of two systems is shown in the inset.
        Color scheme: C (gray), H (white), N (light blue), O (red), Si (blue), B (green), Cl (yellow).
        }}
        \label{fig:barr}
    \end{figure}

    \myred{Our second test set consists of reaction barrier heights ($E_{\mathrm{BH}}$) which govern the rates of chemical reactions and therefore serve as key figures of merit in catalysis.
    For this purpose, we choose the BHDIV10 dataset, which is a subset of GMTKN55~\cite{Goerigk17PCCP}.
    The BHDIV10 set contains 10 chemically diverse transition states (two of them are visualized in the inset of \cref{fig:barr}), with $E_{\text{BH}}$ spanning a wide range from $10$ to $100$~kcal/mol.
    We performed LNO-CCSD and LNO-CCSD(T) calculations by tightening $\eta_{\mathrm{vir}}$ from $3\times 10^{-5}$ to $3\times 10^{-7}$ for all 10 systems in the BHDIV10 set using aug-cc-pVTZ basis sets and various LNO constructions and external corrections.
    In \cref{fig:barr}, we present the convergence of the LNO-CCSD and LNO-CCSD(T) barrier heights toward their respective canonical references in terms of the mean absolute error averaged over all 10 reactions as a function of the average maximum LAS size.}

    \myred{When external corrections are not included, RPA-based LNO-CCSD and LNO-CCSD(T) show accuracies similar to those of their MP2-based counterparts at comparable LAS sizes.
    This behavior is fully consistent with the results for noncovalent molecules in \cref{subsec:non_covalent_molecules} and further confirms the comparable quality of the RPA- and MP2-based LNO constructions.
    When external corrections are included, the fastest convergence is obtained with SOSEX-based corrections for LNO-CCSD and with MP2-based corrections for LNO-CCSD(T), again in good agreement with our observations for noncovalent molecules.
    Noticeable differences, however, are observed between the RPA- and SOSEX-based corrections: relative to SOSEX, the RPA-based correction leads to slightly slower convergence for LNO-CCSD but faster convergence for LNO-CCSD(T).
    This difference is consistent with the greater importance of short-range exchange in barrier heights than in noncovalent interactions, which amplifies the distinction between RPA and SOSEX for this test set.
    Nonetheless, all variants converge robustly to within chemical accuracy of the canonical reference already with an average LAS size of fewer than $100$ active orbitals, which corresponds to a relatively loose threshold of $\eta_{\mathrm{vir}} = 10^{-5}$.}

    \subsection{Bulk metals}
    \label{subsec:bulk_metals}

    The third class of materials used to assess the performance of RPA-based LNO-CC consists of bulk metals.
    In these systems, the vanishing band gap and the finite density of states at the Fermi level render MP2 and, more generally, any finite-order perturbation theory formally divergent in the TDL~\cite{Shepherd13PRL,Masios23PRL,Neufeld23PRL}.
    In practice, however, calculations are always performed in finite simulation cells with discrete $k$-point sampling, which introduce an effective finite gap and make MP2 a qualitatively meaningful approximation.
    Indeed, numerous previous studies have demonstrated the practical success of using MP2-based natural orbitals, wavefunction amplitudes (e.g.,~through structure-factor analysis~\cite{Mihm21NCS}), and energy corrections to accelerate CC calculations for metallic systems~\cite{Callahan21JCP,Neufeld22JPCL,Ye24JCTC,Carbone24FD,Schafer25JPCL}.
    Despite these successes, it remains unclear whether RPA-based counterparts can offer tangible advantages over existing MP2-based methods.
    This question is the focus of the present section.

    \subsubsection{BCC Li and FCC Cu with fixed $k$-point mesh}
    \label{subsubsec:li_cu}

    \begin{figure}[!h]
        \centering
        \includegraphics[width=3in]{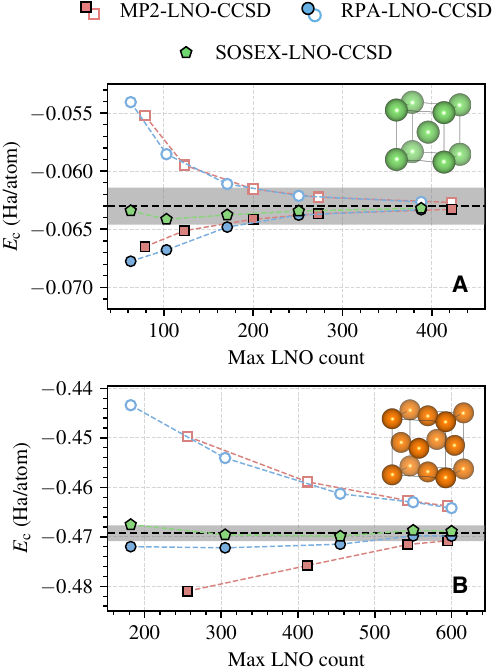}
        \caption{Convergence of MP2-, RPA-, and SOSEX-based LNO-CCSD correlation energies per atom as a function of the maximum LAS size for (A) body-centered cubic (BCC) Li with a $3\times3\times3$ $k$-mesh and a lattice constant of $3.51$~\AA{}, and (B) face-centered cubic (FCC) Cu with a $2\times2\times2$ $k$-mesh and a lattice constant of $3.615$~\AA{}.
        The Li calculations employ the GTH-cc-pVTZ basis sets, while the Cu calculations use the all-electron cc-pVTZ basis sets optimized in this work.
        Open and filled symbols denote LNO-CCSD results obtained without and with the external energy correction, respectively.
        The uncorrected SOSEX-based LNO-CCSD results are identical to the uncorrected RPA-based results and are therefore not shown.
        The best estimate of the canonical CCSD correlation energy, obtained by averaging the best extrapolated values from all three methods using \cref{eq:Ec_extrap_ext}, is indicated by a horizontal dashed line in each panel.
        The gray shaded region represents an error window of $\pm 1$~kcal/mol relative to the corresponding reference value.}
        \label{fig:li_cu_cmp}
    \end{figure}

    In \cref{fig:li_cu_cmp}, we compare the convergence behavior of MP2- and RPA-based LNO-CCSD correlation energies for two representative bulk metals: body-centered cubic (BCC) Li and face-centered cubic (FCC) Cu.
    In earlier work~\cite{Ye24JCTC}, MP2-based LNO-CCSD was successfully applied to investigate the ground-state bulk properties of BCC Li.
    As shown in \cref{fig:li_cu_cmp}A, replacing MP2 with RPA yields very similar convergence behavior of the LNO-CCSD correlation energy, both with and without $\Delta E_{\mathrm{c}}^{\mathrm{ext}}$.
    The inclusion of $\Delta E_{\mathrm{c}}^{\mathrm{ext}}$ reduces the absolute correlation energy error, particularly for small LAS sizes.
    However, this improvement is noticeably less pronounced than that observed for gapped systems (cf.~\cref{fig:c2c2pd_conv}A), consistent with previous findings~\cite{Ye24JCTC}.

    A qualitatively different behavior is observed for FCC Cu, as shown in \cref{fig:li_cu_cmp}B.
    While the uncorrected LNO-CCSD correlation energies obtained from MP2- and RPA-based LNOs remain similar, the inclusion of RPA-based $\Delta E_{\mathrm{c}}^{\mathrm{ext}}$ leads to a substantial reduction in error, achieving an accuracy of approximately $2$~kcal/mol with as few as $200$ active LNOs.
    By contrast, achieving comparable accuracy with the MP2-based external correction requires roughly $500$ active LNOs.
    This contrasting behavior can be rationalized by examining the relative performance of MP2 and RPA for the two metallic systems.
    As summarized in Table~S2, RPA significantly overestimates the CCSD correlation energy for Li by about $36\%$, but only slightly for Cu (about $3\%$).
    In contrast, MP2 fortuitously yields a correlation energy close to the CCSD reference for Li for the $3\times3\times3$ $k$-point mesh employed, underestimating it by only about $4\%$, while substantially overestimating the CCSD reference for Cu by roughly $20\%$.
    These trends directly impact the effectiveness of the corresponding external corrections.

    To further assess the role of beyond-RPA corrections, we also include in \cref{fig:li_cu_cmp} the results for both metals obtained using SOSEX-based LNO-CCSD, which replaces the RPA-based external correction [\cref{eq:Ec_ext_RPA}] with the SOSEX-based one [\cref{eq:Ec_ext_SOSEX}].
    Remarkably, employing the SOSEX-based correction outperforms both MP2- and RPA-based corrections, achieving chemical accuracy for both systems even at the loosest LNO truncation thresholds tested, which corresponds to approximately $50$ active LNOs for Li and fewer than $200$ for Cu.
    Interestingly, despite this strong performance as an external correction, SOSEX itself is not a particularly accurate approximation to the CCSD correlation energy, systematically underestimating it by about $20\%$ for both metals.
    Taken together, these results underscore that the effectiveness of LNO-CC hinges not only on the quality of the LNOs, but also critically on the choice of the external energy correction.

    \subsubsection{BCC Li with enlarging $k$-point meshes}
    \label{subsubsec:li_large_k}

    \begin{figure*}[!t]
        \centering
        \includegraphics[width=6in]{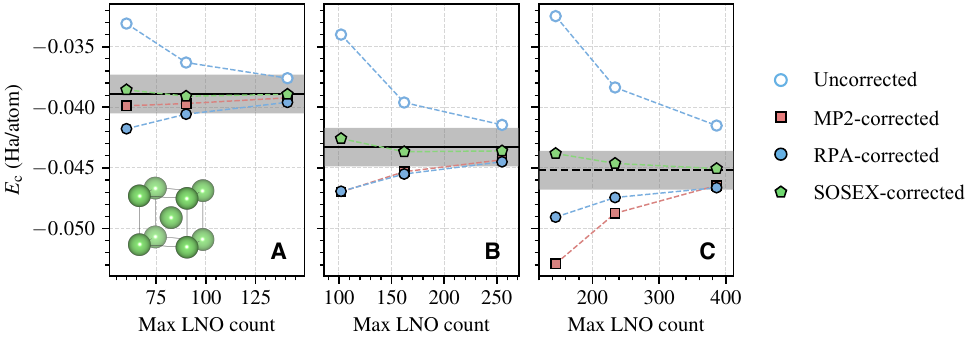}
        \caption{Convergence of LNO-CCSD correlation energies for BCC Li as a function of the maximum LAS size using RPA-derived LNOs with different external corrections $\Delta E_{\mathrm{c}}^{\mathrm{ext}}$, for (A) $2\times2\times2$, (B) $3\times3\times3$, and (C) $4\times4\times4$ $k$-point meshes.
        All $k$-meshes are shifted from the $\Gamma$ point by $\bm{k}^* = \frac{2\pi}{L}(\frac{1}{4},\frac{1}{4},\frac{1}{4})$, where $L = N_k^{1/3} a$ and $a=3.51$~\AA{} is the lattice constant.
        All calculations employ the GTH-cc-pVDZ basis sets.
        In panels (A) and (B), the horizontal solid lines indicate reference values obtained from canonical $k$-point CCSD calculations.
        In panel (C), the horizontal dashed line denotes our best extrapolated estimate obtained using \cref{eq:Ec_extrap_ext} with the SOSEX external correction [\cref{eq:Ec_ext_SOSEX}].
        The gray shaded region in each panel represents an error window of $\pm 1$~kcal/mol relative to the corresponding reference value.
        }
        \label{fig:li_dz_tdl}
    \end{figure*}

    The results discussed in \cref{subsubsec:li_cu} were obtained at fixed $k$-point meshes and may therefore be affected by finite-size errors.
    This issue is particularly relevant for MP2-based treatments, whose accuracy is expected to degrade as the TDL is approached.
    To systematically assess the behavior of different low-level theories with respect to $k$-point convergence, we repeated the LNO-CCSD calculations for BCC Li using progressively denser $k$-point meshes, from $2\times2\times2$ to $4\times4\times4$, employing the smaller GTH-cc-pVDZ basis sets.
    In these calculations, the RPA-derived LNOs are kept fixed, and we focus exclusively on comparing the impact of different choices of the external correction.
    The results obtained for three representative LNO truncation thresholds ($3\times10^{-4}$, $10^{-4}$, and $3\times10^{-5}$) are summarized in \cref{fig:li_dz_tdl}.

    As the $k$-point mesh becomes denser, the error of the uncorrected LNO-CCSD results increases at each truncation threshold, even though the corresponding LAS sizes continue to grow.
    This behavior reflects the slow decay of the long-range correlation energy in metallic systems and highlights the necessity of incorporating an appropriate external correction.
    With the relatively coarse $2\times2\times2$ $k$-mesh, the MP2-based correction yields rapid convergence with the LAS size.
    However, its performance deteriorates markedly as the $k$-mesh is refined.
    For the largest $4\times4\times4$ mesh, the MP2-based correction provides only marginal improvement in the absolute error over the uncorrected results.
    In contrast, both RPA- and SOSEX-based external corrections yield results that are stable across different $k$-point meshes.
    At the $4\times4\times4$ $k$-mesh, the RPA-based external correction clearly outperforms the MP2-based counterpart \myred{at small LAS sizes, while showing similar accuracy to the MP2-based correction for the largest LAS size tested}.
    For all $k$-meshes considered, the RPA-corrected LNO-CCSD attains chemical accuracy when using the tightest threshold tested ($3\times10^{-5}$).
    Remarkably, the SOSEX-corrected LNO-CCSD achieves chemical accuracy even with the loosest threshold ($10^{-4}$), reducing the required LAS size by approximately a factor of three relative to the RPA-based correction.
    The superior performance of the SOSEX-based external correction is consistent with the trends observed earlier in \cref{fig:li_cu_cmp} for fixed $k$-meshes.

    \section{Conclusion}
    \label{sec:conclusion}

    In this work, we have developed RPA and the closely related SOSEX as promising alternatives to MP2 as the low-level theory within the LNO-CC framework by leveraging the drCCD formulation of RPA.
    \myred{Our numerical benchmarks on noncovalent molecular complexes and crystals, reaction barrier heights, and bulk metals demonstrate that RPA- and SOSEX-based LNO-CC closely match the performance of their MP2-based counterpart at the LNO-CCSD(T) level.
    At the LNO-CCSD level, in contrast, external energy corrections based on RPA and SOSEX systematically outperform those based on MP2, yielding markedly faster convergence toward the canonical coupled-cluster limit.
    The improvement is particularly pronounced for metallic systems as the TDL is approached, and is also clearly evident for noncovalent interactions and barrier heights.
    These results identify RPA and SOSEX as compelling alternatives to the commonly used MP2 when constructing practical LNO-CC and related local correlation schemes.}


    Several avenues for future work naturally follow from the present study.
    First, beyond the encouraging numerical results reported here, a deeper theoretical understanding of why certain composite corrections---most notably SOSEX---perform substantially better than others would be highly valuable.
    Second, in addition to SOSEX, other beyond-RPA approaches that incorporate exchange effects~\cite{Hebelmann12PRA,Chen18JCTC} or higher-order contributions~\cite{Hebelmann11JCP,Shepherd14JCP,Cieslinski23JCTC} offer promising opportunities to further enhance the accuracy and robustness of LNO-CC.
    Finally, the RPA-based strategy presented in this work is general and can be straightforwardly extended to other MP2-based fragment embedding and local correlation frameworks.
    With continued methodological developments along these lines, we anticipate that RPA and its extensions will play an increasingly important role in enabling accurate and efficient high-level quantum chemical calculations for complex molecular and condensed-phase systems.

    \section*{Supporting Information}

    See the Supporting Information for (i) detailed basis-set information, (ii) convergence of RPA-dLNO-based LNO-CC calculations for C2C2PD, (iii) convergence of RPA-, MP2-, and SOSEX-based LNO-CC calculations for GGG and the anthracene crystal, (iv) computational cost of RPA-based LNO-CCSD(T) for C2C2PD, and (v) correlation energies of BCC Li and FCC Cu obtained using MP2, RPA, SOSEX, and CCSD.

    \section*{Conflict of interest}
    The authors declare no competing conflicts of interest.

    \section*{Data availability}

    The data that support the findings of this study are available from the corresponding author upon reasonable request.

    \section*{Acknowledgments}

    This work was supported by the National Science Foundation under Grant No.~CHE-2543461.
    We acknowledge computing resources provided by the Division of Information Technology at the University of Maryland, College Park, as well as by the National Science Foundation through the ACCESS program under allocation CHE240210.

    \bibliography{refs}

\end{document}


\title{Supporting Information for Random phase approximation-based local natural orbital coupled cluster theory}

    \author{Ruiheng Song}
    \affiliation{Department of Chemistry and Biochemistry, University of Maryland, College Park, Maryland, 20742}

    \author{Xiliang Gong}
    \affiliation{Department of Chemistry and Biochemistry, University of Maryland, College Park, Maryland, 20742}

    \author{Aamy Bakry}
    \affiliation{Department of Chemistry and Biochemistry, University of Maryland, College Park, Maryland, 20742}

    \author{Hong-Zhou Ye}
    \email{hzye@umd.edu}
    \affiliation{Department of Chemistry and Biochemistry, University of Maryland, College Park, Maryland, 20742}
    \affiliation{Institute for Physical Science and Technology, University of Maryland, College Park, Maryland, 20742}
    \date{\today}

    \maketitle

    \section{Basis sets and geometries}

    The solid-optimized cc-pVTZ basis sets for Cu, as well as the fitting basis sets for Cu and Li used in this work, are available in the following GitHub repository:
    \begin{center}
        \href{https://github.com/hongzhouye/supporting\_data/tree/main/2026/RPALNOCC}{github.com/hongzhouye/supporting\_data/tree/main/2026/RPALNOCC}
    \end{center}
    The same repository also contains the structure files for the C2C2PD molecular dimer and the anthracene molecular crystal, including all geometries required for counterpoise corrections.

    \section{Supplementary figures}

    \begin{figure}[!h]
        \centering
        \includegraphics[width=0.9\linewidth]{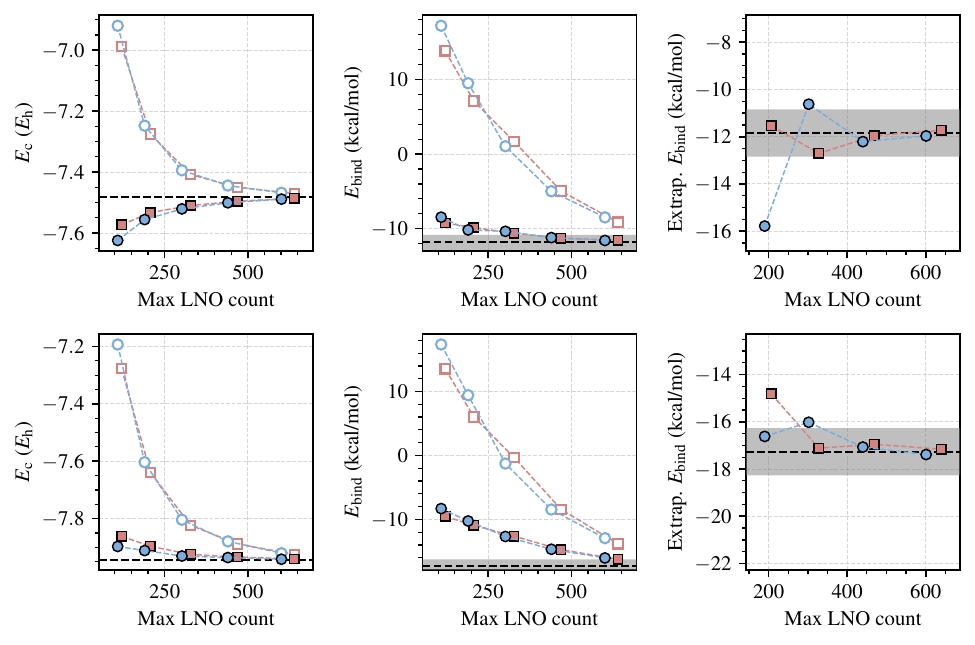}
        \caption{LNO convergence results corresponding to FIG.~\fakeref{2} in the main text, but comparing RPA-based dLNOs (red) with RPA-based LNOs (blue).
        The results for the RPA-based LNOs are identical to those shown in FIG.~\fakeref{2} of the main text.\\
        Alt text: Plots of LNO-CC correlation and binding energies for the C2C2PD complex as a function of LAS size (maximum number of active LNOs), comparing RPA-based LNO-CC using normal and direct LNO definitions.
        }
    \end{figure}

    \begin{figure}[!h]
        \centering
        \includegraphics[width=0.9\linewidth]{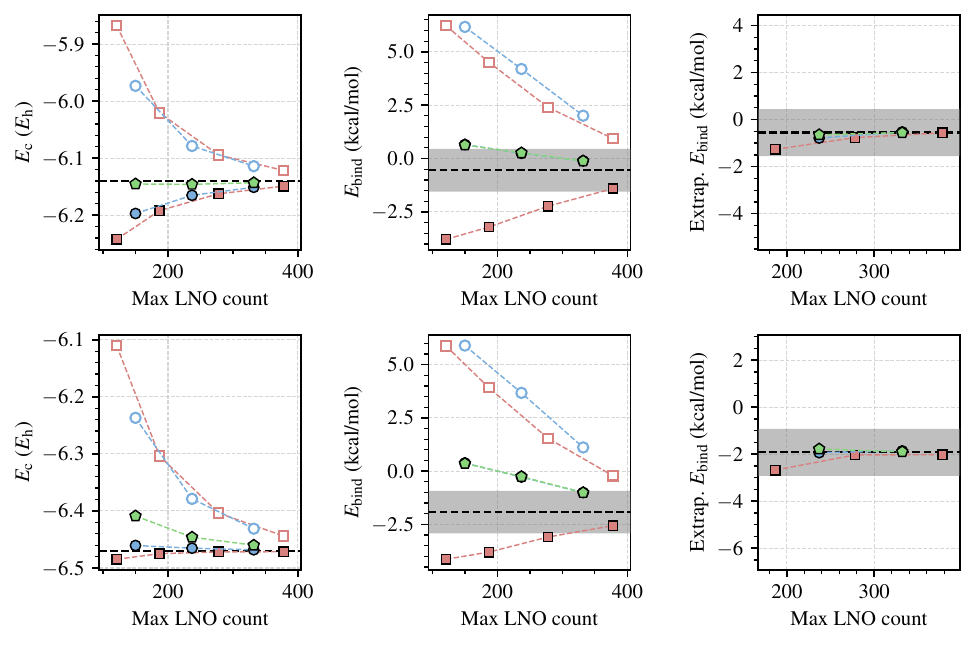}
        \caption{LNO convergence results corresponding to FIG.~\fakeref{2} in the main text, but for GGG in the aug-cc-pVTZ basis sets.
        Red, MP2-LNO-CC; blue, RPA-LNO-CC; green, SOSEX-LNO-CC.\\
        Alt text: Plots of LNO-CC correlation and binding energies for the GGG molecular complex as a function of LAS size (maximum number of active LNOs), comparing MP2-, RPA-, and SOSEX-based approaches.
        }
    \end{figure}

    \begin{figure}[!h]
        \centering
        \includegraphics[width=0.9\linewidth]{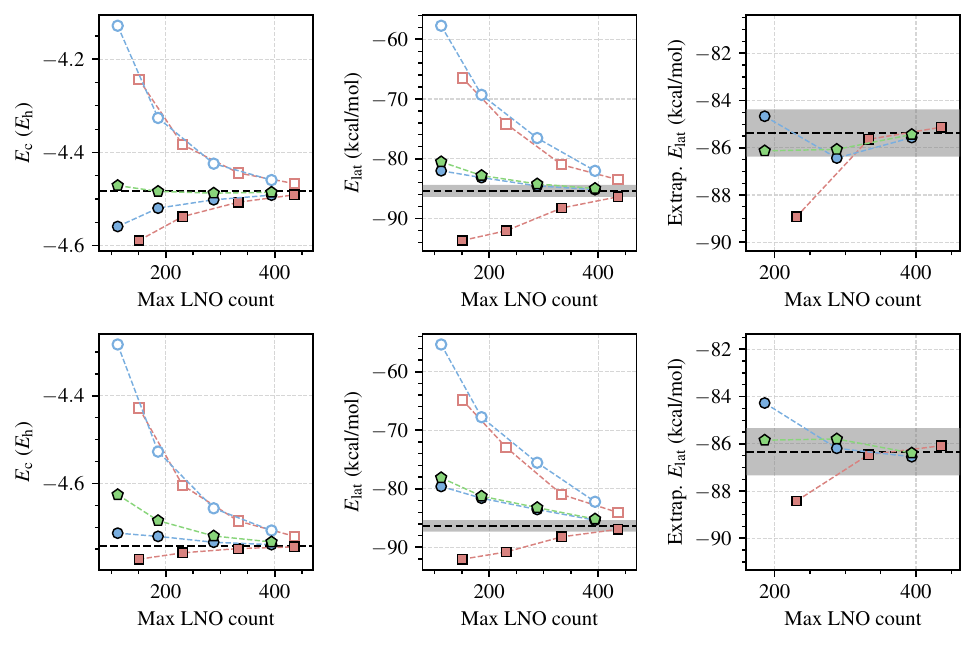}
        \caption{LNO convergence results corresponding to FIG.~\fakeref{2} in the main text, but for the anthracene crystal in the cc-pVTZ basis sets with $\Gamma$-point Brillouin zone sampling.
        Red, MP2-LNO-CC; blue, RPA-LNO-CC; green, SOSEX-LNO-CC.\\
        Alt text: Plots of LNO-CC correlation and binding energies for the anthracene molecular crystal as a function of LAS size (maximum number of active LNOs), comparing MP2-, RPA-, and SOSEX-based approaches.
        }
    \end{figure}


    \clearpage

    \section{Supplementary data}

    \begin{table}[!h]
        \centering
        \caption{Computational cost of the individual steps in an RPA-based LNO-CCSD(T) calculation for C2C2PD using the cc-pVTZ basis sets and an LNO truncation threshold of $\eta_{\mathrm{vir}} = 3 \times 10^{-7}$.
        The drCCD equations are solved only once, whereas the reported wall times for the RPA-based LNO construction and the fragment CC calculations are accumulated over all $108$ fragments of this molecule.
        In practice, the fragment calculations are carried out in parallel, with each fragment assigned 24 EPYC 7763 CPU cores and 96~GB of total memory.\\
        Alt text: Table of LNO-CC cost breakdown by individual steps for the C2C2PD molecular complex, including drCCD equation, RPA LNO construction, fragment CCSD, and fragment (T).
        }
        \begin{tabular}{lc}
            \toprule
                Step & Wall time (hour) \\
            \midrule
                drCCD equation & 1.1 \\
                RPA LNO construction & 5.5 \\
                Fragment CCSD & 943 \\
                Fragment (T) & 3750 \\
            \bottomrule
        \end{tabular}
    \end{table}


    \begin{table}[!h]
        \centering
        \caption{Per-atom correlation energy ($E_{\mathrm{h}}$) calculated by different methods for BCC Li and FCC Cu using different basis sets and $k$-point meshes.
        The CCSD results for $2\times2\times2$/DZ and $3\times3\times3$/DZ are from canonical $k$-point CCSD, while all others are the best extrapolated values from LNO-CCSD as explained in the main text.\\
        Alt text: Table of LNO-CCSD correlation energies (Eh) for BCC lithium and FCC copper across different k-point meshes and basis sets, comparing MP2-, RPA-, and SOSEX-based LNO-CCSD results with canonical CCSD.
        }
        \label{tab:li_cu_cmp}
        \begin{tabular}{lcccccc}
            \toprule
            System & $N_k$ & Basis set & MP2 & RPA & RPA+SOSEX & CCSD    \\
            \midrule
            BCC Li & $3\times3\times3$ & TZ & $-0.0604$ & $-0.0862$ & $-0.0488$ & $-0.0630$  \\
            \midrule
            \multirow{3}*{BCC Li}
            & $2\times2\times2$ & DZ & $-0.0344$ & $-0.0478$ & $-0.0276$ & $-0.0388$  \\
            & $3\times3\times3$ & DZ & $-0.0463$ & $-0.0535$ & $-0.0325$ & $-0.0434$  \\
            & $4\times4\times4$ & DZ & $-0.0534$ & $-0.0557$ & $-0.0346$ & $-0.0452$  \\
            \midrule
            FCC Cu & $2\times2\times2$ & TZ & $-0.5583$ & $-0.4823$ & $-0.3789$ & $-0.4690$  \\
            \bottomrule
        \end{tabular}
    \end{table}
